\documentstyle[psfig]{elsart}

\begin{document}
\maketitle

\begin{frontmatter}

\title{\sc Are You Ready to FLY in the Universe ?  A Multi-platform N-body Tree Code 
for Parallel Supercomputers} 

\author[oact]{U. BECCIANI} and
\author[oact]{V. ANTONUCCIO-DELOGU} 

\address[oact]{Osservatorio Astrofisico di Catania,
Citt\`{a} Universitaria, Via S. Sofia, 78 -- I-95125 Catania - Italy \\
e-mail: ube@sunct.ct.astro.it van@sunct.ct.astro.it}

\begin{abstract}
In the last few years, cosmological simulations of structures and galaxies 
formations have assumed a fundamental role in the study of the origin,  formation and 
evolution of the universe. These studies improved enormously with the use of 
supercomputers and parallel systems, allowing more accurate simulations,  
in comparison with traditional serial systems.
The code we describe, called FLY, is a newly written code (using the tree 
N-body method), for  three-dimensional self-gravitating collisionless systems evolution.\\
FLY is a fully parallel code based on the tree Barnes-Hut algorithm and 
 periodical boundary conditions are implemented by means of the Ewald summation 
technique. We use FLY
to run simulations  of the large scale structure of the
 universe  and   of cluster of galaxies, but it could be usefully adopted 
to run evolutions of systems based on a tree N-body algorithm.
FLY is based on the one-side communication paradigm to share data among the processors, 
that access to remote private data avoiding any kind of synchronism. The code was
originally developed on CRAY T3E system using the logically SHared MEMory access routines ({\it SHMEM})
but it runs also
 on SGI ORIGIN systems and on IBM SP by using the Low-Level Application Programming Interface routines ({\it LAPI}).\\
This new code is the evolution of preliminary codes (WDSH-PT and WD99)
 for cosmological simulations we implemented
in the last years, and it reaches very high performance in all systems where 
it has been well-tested. This performance allows us today to consider the code FLY among the most powerful
parallel codes for tree N-body simulations. The performance that FLY reaches is 
discussed and reported, and a comparison with other similar codes is 
preliminary considered.
The FLY version 1.1 is freely available on http://www.ct.astro.it/fly/ 
 and it will be 
maintained and upgraded with new releases.

\noindent 
\end{abstract}

\end{frontmatter}

\section{\bf Introduction}

Numerical simulations are  very important tools to study the origin and
the evolution of the universe, the cluster of galaxies and  
galaxy formations. They play a fundamental role in testing and verifying
several cosmological theories, adopting different  initial conditions and 
models of the expansion of the universe that affect the formation and
the evolution of the large scale structures (hereafter LSS) \cite{bar86} \cite{her87} \cite{dub88} and the matter property and
distribution. This class of numerical simulations is referred to as the N-body problem class.\\
Observations covering the entire span of electro-magnetic spectrum
allow us to estimate the amount of {\it visible} matter. But only a small fraction of the matter of the 
universe is visible, and it could  be that up to  ninety-five per cent of the matter 
is {\it dark} and does not emit any form of electro-magnetic radiation. It is the dark matter that is 
governing the  dynamics of the universe. The dark matter is modelled as  a self-gravitating collisionless 
fluid, described by the collisionless Boltzmann equation.\\
The method we adopt to run LSS simulations, is a tree based algorithm that places the particles in
 hierarchical groups. 
The fundamental idea of the tree codes consists in  the approximation 
of the force component for a particle. Considering a region $\gamma$, the force component 
on an {\it i-th} particle may be computed as

\begin{equation}
\sum_{j\in\gamma} - \frac{Gm_j {\bf d}_{ij}}{\mid d_{ij}\mid^3} \approx \frac{GM {\bf d}_{i,cm}}
{\mid d_{i,cm}\mid^3} + \; \hbox{higher order multipoles terms }\; 
\label{eq1}
\end{equation}
 where $M = \sum_{j\in\gamma}m_j$ and $cm$ is the  center of mass of $\gamma$.\\
In  Eq. (\ref{eq1}) the multipole expansion is carried out up to the quadrupole order when a {\it far} group is 
considered.
The tree  method,  having no geometrical constraints, adapts dynamically the tree structure to the 
particles distribution  and to the clusters, without loss of accuracy. This method scales as $O(NlogN)$.\\
The most popular tree algorithm for cosmological simulations is the algorithm proposed 
by Barnes and Hut in 1986 \cite{barh86},  including three main
phases. In the first phase, the system is first surrounded by a single cubic region, 
 encompassing all  the particles, that forms
the {\it root-cell} of the tree. The next tree levels are
formed by using the Orthogonal Recursive Bisection (ORB). During the force compute phase (hereafter FC)
an  {\it interaction list} (hereafter IL) is formed for each particle. Starting from the root-cell, and
analysing the tree level by level, the ratio $C_{ellsize}/d_{i-cell}$ 
is compared with an opening angle parameter $\theta$ (generally ranging from 0.5 to 1.0), being
 $d_{i-cell}$  the distance between the particle and the center of mass of the cell. 
If the ratio is smaller than $\theta$ the cell is {\it closed}, added to the IL
and it is 
considered as a  {\it far} region.  The sub-cells of a closed cell will not be   
investigated in the next tree 
level any longer. Otherwise, the cell  is {\it opened} and,
in the next tree level analysis,  the sub-cells will be checked using the same criterion. This procedure
will be repeated 
until all the tree levels are considered. All the particles
found during the tree analysis are always added to the IL. In the last  phase all the particles positions
 are updated, before starting a new cycle.
Each particle evolves following the laws of Newtonian physics. Generally,
in a very large simulation, the differential 
equations are integrated using the numerical Leapfrog integration scheme: 

\begin{equation}
\frac{{\bf x}^{n+1} - {\bf x}^{n}}{\Delta T} = {\bf v}^{n+1/2} 
\label{eq2}    
\end{equation}
\begin{equation}
{\bf v}^{n+1/2} - {\bf v}^{n-1/2} = \frac{{\bf F}^n \Delta T}{m}
\label{eq3}
\end{equation}

\noindent where $\Delta T$ is the discrete time-step and the superscript $n$ refer to the time instant
$t=n \Delta T$.\\
The  parallel code FLY ({\bf F}ast {\bf L}evel-based N-bod{\bf Y} code) we describe, is a tree algorithm  
code to run simulations of collisionless
gravitational systems with a larger number of particles ($ N \geq 
10^7$). FLY incorporates  fully periodic boundary conditions using the Ewald method, without the use of the 
Fast Fourier Transform techniques \cite{her91} and it is designed for MPP/SMP systems using the one-side 
communication paradigm.\\
Section 2  show an overview of the FLY design. We 
will discuss the four main characteristics of FLY in Sections 3, 4, 5 and 6 respectively 
(domain decomposition, 
grouping,  dynamic load balance and  data buffering). Section 7
shows the results of our tests and some comparisons with other codes, Section 8 presents our conclusions.

\section{\bf FLY parallel code}

FLY is the code we design, develop and use to run very big simulations of the LSS of
the universe using parallel systems MPP and/or SMP. It is based on the tree algorithm 
 described above and all the phases are fully parallelized.
FLY uses the Leapfrog numerical integration scheme for performance reasons,  and 
incorporates  fully periodic boundary conditions using the Ewald method, without using  the 
fast Fourier  transform techniques \cite{her91}.\\
FLY is the result of a project started in 1994 in order to produce a code for collisionless
cosmological simulations. We start with the design of a parallel code  for a workstations
cluster \cite{ant94}, based on the locally essential tree \cite{sal90} \cite{sal97} \cite{dub96}, and using the
PVM \cite{pvm} library. The code did not give high performance results due to the low network 
bandwidth and high latency;  moreover, the number of particles that was possible to simulate was small, due to the
big size of the locally essential tree.\\
The choice we did in 1996 was to re-design the code without using a locally essential
tree and using the CRAY T3D with CRAFT \cite{craft}. The new code was called WDSH-PT \cite{bec96} 
(Work and data SHaring - Parallel Tree code) and  a dynamically balanced version 
was produced in 1997 \cite{bec97}.\\
  In 1999, to allow simulations with larger 
resolution, we  first re-considered the 
grouping strategy as described in J. Barnes (1990) \cite{bar90}
and applied it, with some modifications, to our WDSH-PT. 
This  code wass called WD99 \cite{bec2000} and the obtained performance in terms of
particles/second was very high.\\
FLY is based on WD99 and is designed for MPP and SMP systems. It is written in Fortran 90 and 
uses the one-side communication paradigm; it has been developed on the CRAY T3E using the SHMEM library. 
FLY mainly uses  remote GET and PUT operations and some atomic operations of global counters. The one-side 
communication paradigm avoids the processors synchronism during all the phases of the tree algorithm. 
This choice gives  FLY an obvious increment of 
performances, moreover this paradigm allows us to port FLY in all platforms where the one-side 
communication is available.\\
FLY is based on four main characteristics described in the following sections. It adopts a simple
 domain decomposition, a grouping strategy,
a dynamic load balance mechanism without overhead, and a data buffering that allows us to
minimize data communication.\\
Data input and data output contain positions and velocities of all particles 
and are written without any {\it control words}. The data format is integrated
with a package that we develop for data analysis: ASTROMD \cite{amd2000}, a freely available software 
(http://www.cineca.it/astromd) for collisionless and gas-dynamical cosmological simulations.\\
The FLY version 1.1, described in this paper, is freely available. It runs on CRAY T3E, SGI ORIGIN 2000 
using the SHMEM library, and on IBM SP using the LAPI library.

\section{\bf Domain decomposition}

Data distribution plays a fundamental role in obtaining a high 
performance of the N-Body codes designed for MPP and SMP systems, and
the domain decomposition is an extremely crucial aspect. Optimal data 
distribution among the processor elements (hereafter PEs), must
avoid any imbalance of the load among the PEs and  minimize the communication on the network. 
Many kinds of codes use a domain decomposition based on splitting planes that subdivide the
domain in sub-domains having the same load, taking into account the mass density distribution. 
The splitting planes produce a final domain decomposition with sub-domains that do not have an equal
geometry;  during the system evolution, to avoid a load imbalance, 
the domain decomposition must be repeated many times.\\
FLY does not split the domain with orthogonal planes, but the domain decomposition is done by assigning an equal number of particles to 
each processor. The data structures of both particles and tree, are subdivided 
among the PEs to ensure a good initial balance of the load and to avoid any bottleneck while accessing remote data.

\subsection{\it Particles data sorting and distribution}

The data input of FLY is the array of  the fields of position and velocity.
Each particle has a tag number from $1$ to $Nbodies$ (the total number of particles).  FLY\_sort, 
an internal utility of FLY, organizes
 the input data by using the same tree domain decomposition of the algorithm. 
Fixing a level $l$ of the tree, FLY\_sort builds the tree up to this level,
where there are $2^{3 \cdot l}$  cells,  $l=0$ being the level of 
the tree root. Then FLY\_sort assigns the 
{\it cell parent} to each particle, the parent being the cell where the 
particle is physically located. At the end, following the tree scheme, 
{\it cell parent by cell parent} and considering in turn the nearest cells, FLY\_sort 
assigns the tag number and stores
all the particles.\\
It is important to choose the $l$ level to fit the number of PEs; the number
of {\it cell parents} must be equal to or greater than the number of PEs: i.e. if the user uses
64 PEs he must fix $l \geq 2$ so as to have $ 2^6 = 64 $ cells.
The final result is a sorted file containing the fields of position and velocity, so that particles with a near 
tag number are also near in the physical space.\\
Data  organized in this way are the input of the FLY simulation code. Each 
processor has the same number of particles, $Nbodies/N\_PES$, near in the space,
 in the reserved and/or local memory, being  $N\_PES$ the number of processors.
 This kind of distribution, in contiguous blocks,
 was already studied in the  WDSH\_PT code
\cite{bec96} and it is the best data distribution in terms of measured code performance.
When  necessary the FLY\_sort procedure may be re-executed, to preserve these properties 
during the system evolution. FLY\_sort consumes a negligible CPU time compared to the
CPU time for a complete simulation run.

\subsection{\it Tree data distribution }

The tree cells are numbered progressively from the root, which encompasses the
whole system,  down to the smallest cells which enclose smaller and smaller regions of
space. The optimal data distribution scheme of the arrays containing the tree properties, both
geometric and physical characteristics, is reached using 
a fine grain data distribution. The first 
tree levels contain cells that are
typically  large enough to contain many particles and, 
during  the simulation,  these cells are checked to form the IL of each particle.
A fine grain size distribution prevents  these cells from being  located in the
same PE, or in a small number of PEs. In fact in  case of a coarse grain size
data distribution,  all the PEs will attempt to access the same PEs memory, 
with the typical
problems of {\it access to a critical resource}. This effect would produce 
a bottleneck that drastically decreases the code performance. On 
the contrary, a tree fine grain data distribution allows, on
average, all the PEs memories to be requested with the same frequency; thus each particle 
will have the same average access time to the tree cells  avoiding the bottleneck problem.\\
We do not claim that this is the optimal choice for mapping the tree onto the T3E
torus or onto other SMP systems, nevertheless the tree data distribution adopted
by FLY gives the best results on all those systems where FLY runs.

\section{\bf The grouping  }

FLY uses the grouping strategy we adopted with  our WD99 code. The basic idea
 is to build a single interaction list
to be applied to all particles inside a {\it grouping cell} $C_{group}$ of the tree.
 This reduces the number of the tree accesses to build the ILs. 
We consider a hypothetical particle we call Virtual Body (hereafter VB)
placed in the center of mass of the $C_{group}$.\\
Using a threshold value R equal to 3 times the $C_{group}$ size 
to limit the errors, the 
interaction list $IL_{VB}$ is formed by two parts:

\begin{equation} 
IL_{VB} = IL_{far} + IL_{near}
\label{eq4}
\end{equation}

\noindent where $IL_{far}$ includes the elements more distant than R from VB  
and $IL_{near}$ includes the elements near  VB. Moreover, all 
$p \in C_{group}$ are included in $IL_{near}$.
Using the two lists in Eq. (\ref{eq4}) it is possible to compute the  force ${\bf F}_p$ 
as the sum of two components:

\begin{equation} 
{\bf F}_p ={\bf F}_{far} +{\bf F}_{near}
\label{eq5}
\end{equation}

The  component  ${\bf F}_{far}$ is computed for  VB,
  using  the elements listed in 
$L_{far}$, and it 
is applied to all the particles $p \in C_{group}$, 
while the  ${\bf F}_{near}$ component is computed separately for each  
particle  with the elements listed in 
$IL_{near}$.  The list  $IL_{near}$ contains only a few elements compared to 
the $IL_{far}$  list, and we obtain  a net gain in performance.
The size of the $C_{group}$ is constrained by the maximum allowed 
value of the overall error of this method. In a 16-million-particle simulation with a box size of 50 Mpc, 
using a conservative level 7 for the size of the $C_{group}$  where the error is lower than 1\%, the
number of the computed particle/second increases by a  factor of 7. In this sense
the performance of FLY is level-based and the fast execution of a time-step  depends on the fixed level for the grouping.

\section{\bf Dynamic Load Balance}

FLY uses the DLB system already used by the WD99 code. Each particle or grouping
 cell has a PE executor (hereafter PEx) that  performs the FC phase for it. 
At  first FLY computes the FC phase for all the grouping cells, 
the default PEx being the processor where the greatest number of  particles
belonging to the $C_{group}$, are memory located. 
When a PE has no more
$C_{group}$ cells to compute, it can start the FC phase for other $C_{group}$ 
cells, not yet
computed by the  default  PEx. The one-side communication paradigm allows FLY to perform
this task without synchronism or waiting states among the PEs, and to get a 
load balance for the FC phase.\\
In a very similar way, FLY balances the load for particles that are not included in grouping cells.
There is a fixed portion $N_{ass}$ of  local particles that must be computed by the local PE
for performance reasons, but the remaining portion $N_{free}$  does not have an
assigned PEx. When each PE completes the FC phase for the $N_{ass}$ particles, it can compute the $N_{free}$
particles of all the PEs. As above, the one-side communication paradigm avoids synchronism or waiting 
states among the PEs, and allows FLY to have a load balance also in the above mentioned FC phase.\\ 
The portion $N_{ass}$ is fixed by the user and, in order to obtain the best performance,
it must be as  large as possible,  thus  each PE can work mainly on the local particles but
it must always work during the FC phase, avoiding the
load imbalance. At the start of each time-step FLY uses, as a prevision value, 
the load of the last time-step, and it
automatically computes the $N_{ass}$ portion again
to guarantee the best performance. The tests we perform show that the best results are
obtained when the $N_{ass}$ quantity ranges from 80\% to 95\% of all particles.

\section{\bf Data buffering}

To compute the FC phase each PEx, for each particle, must check  the
tree cells, level by level, starting from the root cell, to form the IL. 
The IL includes about  90\% of cells 
and, as the cells are distributed among the PEs, each PEx must execute a high 
number of remote accesses. Moreover, generally, a PEx executes the FC phase for near
particles, residing in the local memory, and having a very similar IL. Then it must often 
 check the same tree cells for many particles.\\
The figures reported in Tab. 1 are measured both in uniform (redshift $Z=50$) and clustered conditions 
(redshift $Z=0.3$) for 2, 8 and 16 million  particles, in a region of 50 Mpc with  $\theta=0.8$. 

\begin{center}
\begin{tabular} {|l|l|l|l|l|l|} \hline

	        &  A 	&   B	& C \\  \cline{1-4}
2Ml  uniform   &	1.0 million	&	315 		&  341  \\ 
2Ml  clustered &	1.2 million	&	340 		&  376  \\ 
8Ml  uniform   &	3.9 million	&	355 		&  386  \\ 
8Ml  clustered &	4.8 million	&	390 		&  420  \\ 
16Ml uniform   &	7.8 million	&	375 		&  403  \\ 
16Ml clustered &	10.0 million	&	420 		&  456  \\ \hline
\end{tabular}
\end{center}
\vspace{0.5cm}

{\bf Tab. 1}: A column: number of internal tree cells. B column: average length of the IL. C column: average tree cells checked to form one IL.\\

\vspace{1cm}

In a 16-million-particle  simulation, in clustered conditions, the 
tree has about 10.0 million  cells, the average IL length has 420 elements, 
but a PEx must  checks 456 tree  cells to form one IL. 
Each tree cell access, retrieves 3 positions (8 bytes each), and for
each opened cell it is necessary to retrieves  8 sub-pointers (4 bytes each), whereas for each closed cell, 
it is necessary to retrieve the mass (8 bytes) and the 5-quadrupole momentum (8 bytes each). 
Considering the opened cells, there are 

\begin{equation}
(16 \cdot 10^6 \cdot (456 - 420) \cdot 2 + 16 \cdot 10^6 \cdot 420  \cdot 3 ) \cdot \frac{NPEs - 1}{NPEs} 
\label{eq6}
\end{equation}
\noindent number of remote GETs of contiguous elements, being $NPEs$ the number of PEs used  to run the simulation, and

\begin{equation}
(16 \cdot 10^6 \cdot (456 - 420) \cdot (3 \cdot 8 + 8 \cdot 4) + 16 \cdot 10^6 \cdot 420 \cdot 9 \cdot 8 ) \cdot \frac{NPEs - 1}{NPEs} 
\label{eq7}
\end{equation}

\noindent the total remote data transfer in bytes, to execute the FC phase.  

The number of remote GETs as reported in expression (\ref{eq6}), the latency time of each remote access and
the bandwidth make it difficult to run a big simulation, even with powerful parallel systems having a high number
of CPUs.\\
FLY introduces the {\it data buffering} to limit the number of remote GETs and the global data transfer,
thus obtaining a great improvement in the code performance and scalability. The data buffering uses all the free memory, 
not allocated to store the arrays of particles and the tree cells properties.\\
At the start, FLY statically allocates all the data structures. The
memory occupancy of the code is about 5 Mbytes plus 220 bytes for
each particle. In conservative mode, FLY allocates a tree having a number of cells equal to the number of particles.
In a 16-million-particle simulation with 64 PEs each of them have 250,000 particles, and the local memory occupancy is 
about 68 Mbytes. Using a system with 256 Mbytes of local memory  for each PE,  there is a large quantity of memory 
that can be used
to store the remote data: FLY checks the free space and dynamically allocates arrays, in order to store positions
masses, pointers and quadrupole momenta of each remote tree cell, that the PE investigates during the FC phase.\\
The data buffer is managed with a policy of a {\it simulated cache} in the RAM. The cache arrays have an index array, 
and the mapping of each  element
has a one-to-one correspondence: each remote element has only one line of the simulated local
cache array 
where it can be loaded. Every time the PE has to access a remote element, at first it looks for the local simulated cache and,
if the element is not found, the PE executes the GET calls to down-load the remote element and stores
it in the cache arrays. In this phase FLY computes only the acceleration components for each
particle  so that any problem can arise on the data validity in the simulated cache.\\
In a simulation with 16 million particles clustered, with 32 PEs and 256 Mbytes of local memory, 
without the use of the simulated cache, the PEs execute about $2.1 \cdot 10^{10}$ remote GETs. This value, using 
the data buffering,
decreases at $1.6 \cdot 10^8$ remote GETs, with an enormous advantage in terms of scalability and 
performance.

\section  {\bf FLY performances and scalability}

In this section we show the measured FLY performance in terms of particles/second and the code
scalability using 2,097,152, 8,388,608 and 16,777,216 particles both in uniform and clustered conditions in
a box region of 50 Mpc and $\theta=0.8$. We use a conservative grouping level and the data buffering.
The tests are executed on all the systems where FLY runs, as described in the following sub-sections.
We run two kinds of tests. The first is to measure  the performance and the second is to 
measure  the code scalability. All the measures are executed using dedicated systems and processors.

\subsection{\it FLY  on CRAY T3E}

We use the CRAY T3E/1200e  available at  Cineca (Casalecchio di Reno - Bologna). 
The CRAY T3E is  a physically distributed memory but globally accessible through one-way communication (SHMEM library).
The system has 256 processors  DEC Alfa 21164A, 600 MHz and  308 Gflop/second peaks (1.2 Gflop/second 
for each PE). The network topology is 3D-torus  with 600 Mbytes/second  and 1 -128 80 microseconds latency time.
The global memory is 48 GBytes. There are two sub-pools: 128 PEs with 128 Mbytes Ram, and 128 PEs with 256 Mbytes 
Ram. The global space disk is 200 GBytes. We use the pool with processors having 256 Mbytes to test the FLY performances 
and to obtain 
the highest gains from the data buffer.\\
Fig. 1 shows the code performance obtained by running  simulations with 32 and 64 PEs.
FLY scalability is shown in Fig. 2 considering the case of 16,777,216 particles, where a speed-up factor of 118
is reached using 128 PEs.\\
\begin{figure}
\psfig{file=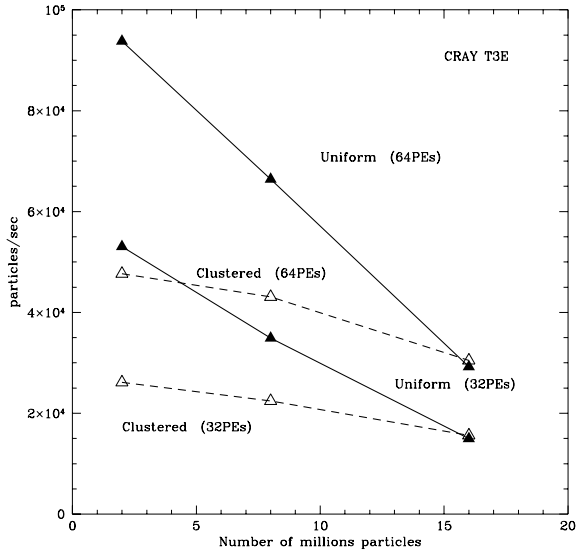,width=8cm}
\caption[h]{CRAY T3E/1200e. FLY particles/second using 32 and 64 PEs in uniform and clustered conditions}
\end{figure}
\begin{figure}
\psfig{file=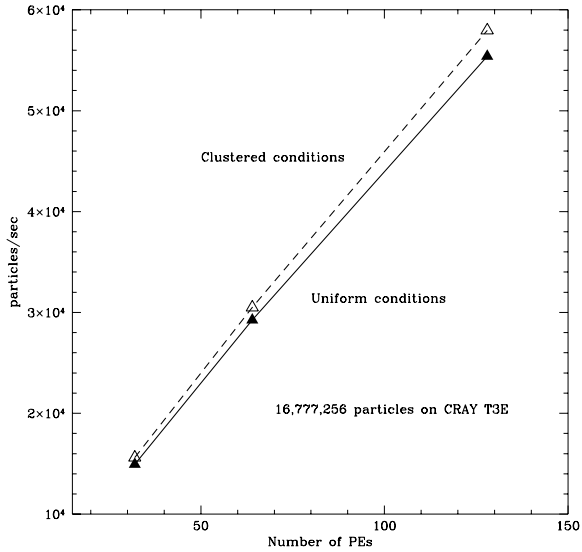,width=8cm}
\caption[h]{CRAY T3E/1200e. FLY scalability from  32 PEs to 128 PEs in a 16-million-particles simulation}
\end{figure}
\noindent The highest performance obtained in a clustered configuration is a positive effect of the grouping characteristic as already
 discussed in WD99.
The obtained results show that FLY has a very good scalability and a very high performance and it
can be used to run very big simulations.

\subsection{\it FLY on SGI ORIGIN 2000}

The system is available at  Cineca. It has 32 nodes and 64 processors MIPS 
Superscalar R12000, 300 MHz, with more than  38 Gflop/second peaks (600 Mflops/second for each PE). Each node has 2 PEs 
and 1 Gbyte memory. The global memory is 32 Gbytes, and the global space disk is 650 Gbytes.  
The ORIGIN 2000 is a CC-NUMA system with globally addressable memory, and distributed shared memory. The interconnecting 
network has 780 Mbytes/second bi-directional transfer rate.\\ 
Fig. 3 shows the code performance obtained by running   simulations with 16 and 32 PEs. The FLY scalability 
 is shown in Fig. 4.\\ 
\begin{figure}
\psfig{file=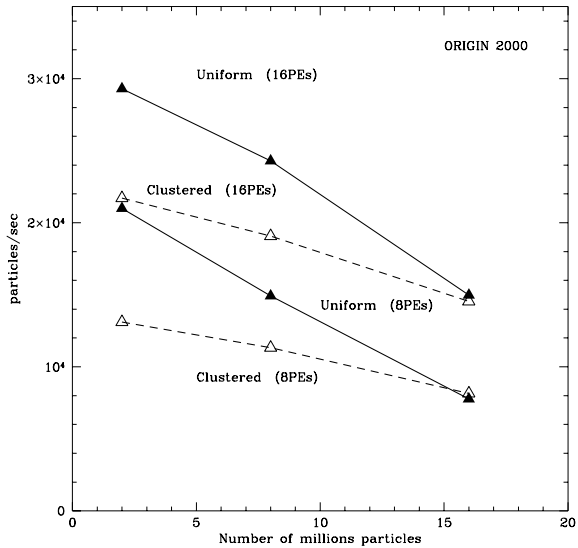,width=8cm}
\caption[h]{SGI ORIGIN 2000. FLY particles/second using 16 and 32 PEs in uniform and clustered conditions}
\end{figure}
\begin{figure}
\psfig{file=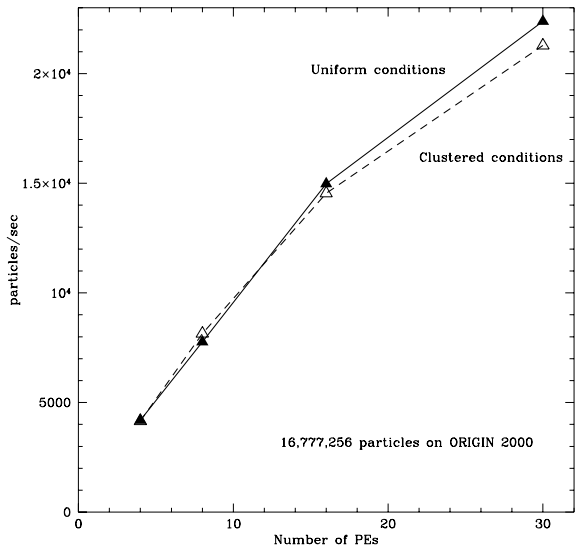,width=8cm}
\caption[h]{SGI ORIGIN 2000. FLY scalability from  4 PEs to 32 PEs in a 16-million-particles simulation}
\end{figure}
\noindent The obtained results show that FLY has still good performance but a lower scalability when more than 16 PEs are used.
 However, FLY can be usefully adopted to run big simulations on this system with a performance comparable with the run on
 the CRAY T3E.

\subsection{\it FLY on IBM SP}

The system is available at the Osservatorio Astrofisico di Catania. It is a distributed memory computer
having 3 nodes and 24  Power 3 RISC superscalar processors, 220 MHz, with more than  20 Gflops/second peaks 
(880 Mflops/second for each PE). Each node has 8 PEs 
and 16 GByte memory. The global memory is 48 GBytes, and the global space disk is 250 GBytes.  The network topology is 
based on an SPS scalable Omega switch having a 300 Mbytes/second bi-directional  transfer rate.\\
FLY uses the LAPI library to perform  one-side communications. The code performance obtained by running simulations 
with 2, 4, 8 and 12 PEs is shown in Fig. 5. FLY scalability is shown in Fig. 6. Unfortunately, running a
parallel code using LAPI, the Omega switch allows the user to use no more than 4 PEs for each node. Moreover,
FLY is optimized when a number of PEs equal to a power of two is used.\\
\begin{figure}
\psfig{file=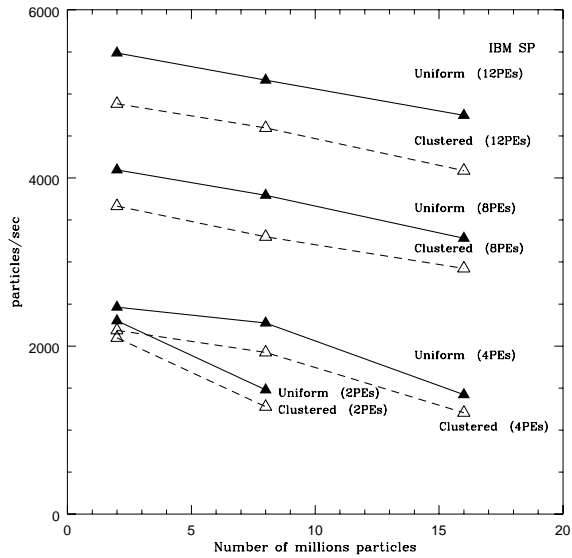,width=8cm}
\caption[h]{IBM SP. FLY particles/second using 2,4,8 and 12 PEs in uniform and clustered conditions}
\end{figure}
\begin{figure}
\psfig{file=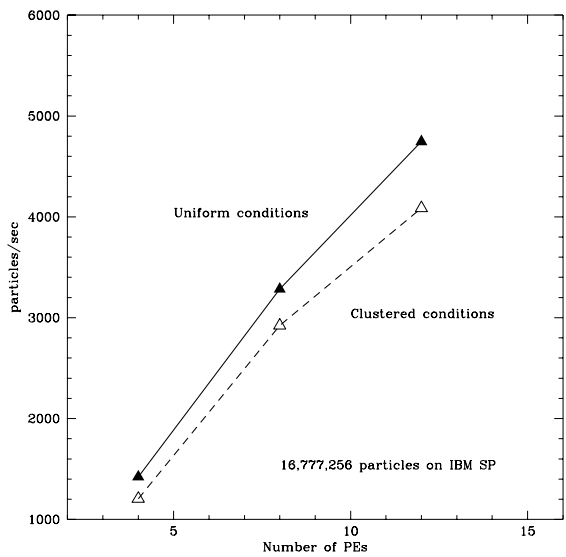,width=8cm}
\caption[h]{IBM SP. FLY scalability from  4 PEs to 12 PEs in a 16-million-particles simulation}
\end{figure}
\noindent The obtained results show that FLY has not such a good performance  and  scalability compared to
 the CRAY T3E system. Therefore, we are developing a  new version of FLY that will use MPI-2, and a new algorithm
 to build the tree. In fact, although on the CRAY T3E system the tree formation phase is shorter than
 5\% of each time-step, this phase takes more than 30\% on the ORIGIN 2000 and on IBM SP, and
 its scalability  is not as good as for the FC phase.

\subsection{\it FLY  comparison}

Fig. 7 shows the FLY performance in all the above systems. We run a 2,097,152 clustered particle simulation 
starting from 2 PEs up to 32 PEs, where availables. The results show that the IBM SP system has lower 
performance than the CRAY T3E and ORIGIN 2000 systems; however, FLY, on CRAY T3E, has a linear scalability 
up to 32 PEs (and over), and reaches a speedup higher than  82 using 128 PEs. The FLY starting performance (4 PEs) 
with the ORIGIN 2000 system is better than with other systems, but the scalability is not as good as with the  CRAY T3E
system.\\
\begin{figure}
\psfig{file=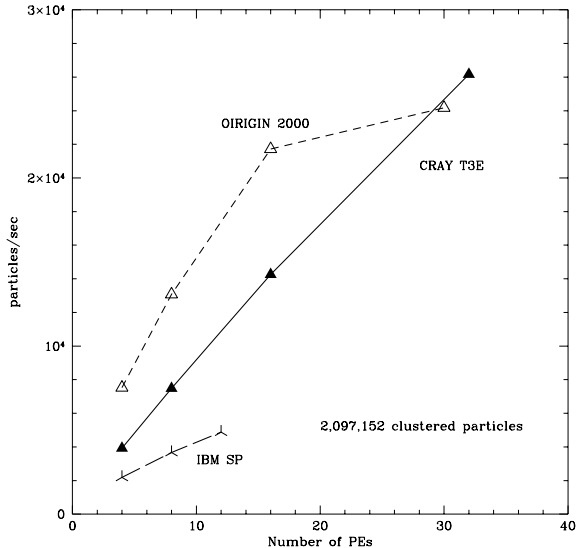,width=8cm}
\caption[h]{FLY performance running 2,097,152 clustered particle simulation on CRAY T3E, ORIGIN 2000 and IBM SP systems.}
\end{figure}
\noindent The results that we obtained can be compared with other similar codes. We consider GADGET \cite{gad2000}, one of the most
 recently released
 tree-SPH codes. Gadget is a code for collisionless and gasdynamical cosmological simulations. GADGET
 implements individual particles timesteps and uses only standard C and standard MPI. It is available on CRAY T3E,
on IBM SP and on Linux-PC clusters, but other platforms could  run the code. With GADGET each processor has a physical
spatial domain assigned and  builds a local tree. The PE can provide the force exerted by its particles for 
any location in 
space. The force computation on a particle therefore requires the communication of the particle coordinates 
to all processors that will 
reply with the partial force components. The total force is obtained by summing up the incoming contributions.\\
Fig. 8 shows a comparison between FLY and data reported by the GADGET authors, considering only the
gravitational section, that is the   heaviest section of GADGET.\\ 
\begin{figure}
\psfig{file=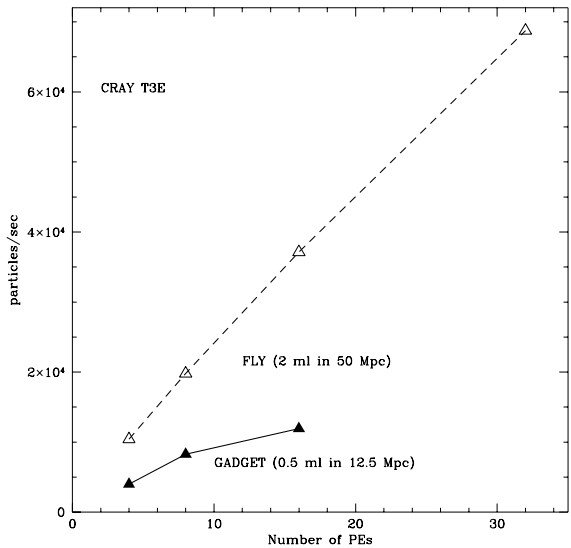,width=8cm}
\caption[h]{FLY and GADGET. A comparison between code performances. The results are obtained using the CRAY T3E system.}
\end{figure}
\noindent FLY seems to have performances higher than  GADGET with the same number
of PEs.  FLY has an optimal
scalability, due to the use of the data buffering. The scalability and speed-up data of the gravitational section 
of GADGET, increasing the number of simulated particles and  the number of PEs to more than 16, are not given 
by the authors. But the GADGET performance, reported in  Fig. 8, has a slope lower than FLY's,  thus a lower 
 scalability, with more than 16 PEs, could  be expected.

\section  {\bf Conclusion and future}

The use of FLY on multiple platforms and the performance obtained enable us  to run simulations with high 
accuracy on the most popular super-computers. FLY is currently used by our research group to run simulations with more than 
16 million particles \cite{ant2000}.
In this sense FLY contributes to study the origin and the evolution
of the universe and allows the execution of simulations of gravitational effects, even if the user has a limited
budget of CPU resources. FLY version 1.1 is freely available and open to the user contribution to enhance the FLY 
code capability, and the porting of FLY on other platforms. The code is written in Fortran 90 and a C language version is
 already in progress. We also plan to use the MPI-2 syntax for a future version of FLY.\\
FLY is used to run simulations considering only the gravitational effects. Even if we are opened to include
the hydrodynamic part in a future version, we plan to integrate FLY with 
other freely available softwares to consider the hydrodynamic effects, including other kinds
of particles (the gas), to study the star formation and  other hydrodynamical effects.

\section{Acknowledgements}

All the tests carried out on the CRAY T3E system and the 
SGI ORIGIN 2000 system at the CINECA, were executed using the financial support
 of the Italian Consortium CNAA (Consorzio Nazionale per l'Astronomia e l'Astrofisica). 
 We gratefully acknowledge useful discussions with Dr. G. Erbacci of CINECA and Dr. C. Magherini of the Department of 
 Mathematics, Florence, for the FLY porting on IBM SP, and Dr. A. F. Lanza of Catania Astrophysical Observatory for 
his useful help.

{}

\end{document}